\documentclass[12pt,preprint]{aastex}

\def\gtsim{\ {\raise-0.5ex\hbox{$\buildrel>\over\sim$}}\ }
\def\ltsim{\ {\raise-0.5ex\hbox{$\buildrel<\over\sim$}}\ }

\begin{document}

\title{
Millimagnitude Photometry for Transiting Extrasolar Planetary Candidates III:
Accurate Radius and Period for OGLE-TR-111-b 
\footnote{Based on observations collected with the 
Very Large Telescope at Paranal Observatory (ESO Programme
075.C-0427(A), JMF and DM visiting observers).
}}

\author{
 Dante Minniti\altaffilmark{1},
 Jos\'e Miguel Fern\'andez\altaffilmark{1,6},
 Rodrigo F. D\'{\i}az\altaffilmark{3},
 Andrzej Udalski\altaffilmark{4}, 
 Grzegorz Pietrzynski\altaffilmark{2,4},
 Wolfgang Gieren\altaffilmark{2},
 Patricio Rojo\altaffilmark{5},
 Mar\'ia Teresa Ru\'iz\altaffilmark{5},
 Manuela Zoccali\altaffilmark{1}
}

\altaffiltext{1}{Department of Astronomy, Pontificia Universidad Cat\'olica, 
Casilla 306, Santiago 22, Chile E-mail:  dante@astro.puc.cl, jfernand@astro.puc.cl, mzoccali@astro.puc.cl}

\altaffiltext{2}{Department of Physics, Universidad de Concepci\'on, Casilla 160-C, Concepci\'on, Chile E-mail: pietrzyn@hubble.cfm.udec.cl, wgieren@astro-udec.cl}

\altaffiltext{3}{Instituto de Astronom\'{\i}a y F\'{\i}sica del Espacio, CONICET- Universidad de Buenos Aires, ArgentinaE-mail: rodrigo@iafe.uba.ar}

\altaffiltext{4}{Warsaw University Observatory, Al. Ujazdowskie 4, 00-478 Waszawa, PolandE-mail: udalski@astrouw.edu.pl}

\altaffiltext{5}{Department of Astronomy, Universidad de Chile, Santiago, Chile
E-mail: mtruiz@das.uchile.cl}

\altaffiltext{6}{Harvard-Smithsonian Center for Astrophysics, Harvard, USA
E-mail: jfernand@cfa.harvard.edu}

\begin{abstract}

We present accurate $V$-band photometry for a planetary transit of OGLE-TR-111
acquired with VIMOS  at the ESO Very Large Telescope.
The measurement of this transit allows to refine 
the planetary radius, obtaining $R_p=1.01 \pm 0.06 ~R_J$.
Given the mass of $M_p = 0.53 ~M_J$ previously measured from radial velocities, 
we confirm that the density is $\rho_p=0.6\pm 0.2 ~g/cm^3$.
We also revise the ephemeris for OGLE-TR-111-b, obtaining an accurate orbital
period $P=4.014484\pm 0.000014 ~$ days, and predicting that
the next observable transits would occur around December 2006, and
after that only in mid-2008. Even though this period is different from previously
published values, we cannot yet rule out a constant period.

\end{abstract}


\section{Introduction}
The discovery of hot Jupiters that transit in front of their parent stars
has advanced our knowledge of extrasolar planets adding a fundamental
datum: the planetary radius.
There has been considerable activity revising the measured radii, owing to
uncertainties in the differential image analysis (see Pont et al. 2006).
It is important to obtain accurate radii from photometry, in order to compare these
exoplanets with the giant planets of the Solar system, and with the
models. In addition, if accurate photometry of transits is available,
one can use timing for future studies of multiplicity in these systems
(e.g. Sartoretti \& Schneider 1999, Miralda-Escude 2002, 
Holman \& Murray 2005, Agol et al. 2005). 

New samples of transiting hot Jupiters
should become available soon (see for example
Fischer et al. 2005, Sahu et al. 2006), but up to now
the OGLE search has provided the largest number of transiting candidates.
In particular,
Udalski et al. (2002) discovered very low amplitude transits in the
$V=16.96$, $I=15.55$ magnitude  star OGLE-TR-111, located in the Carina region
of the Milky Way disk, at $RA(2000)=10:53:17.91$, $DEC(2000)=-61:24:20.3$. 
They monitored 9 individual transits, measuring an amplitude $A_I=0.019$ mag, 
and a period $P=4.01610$ days.
The period is a near-multiple of a day, therefore, the window for transit 
observations is restricted to a couple of months per year.

The planet OGLE-TR-111-b was discovered by Pont et al. (2004)
with precise velocity measurements. They measured $M_P=0.53 ~M_J$, 
$R_p = 1.0 R_J$, and $a=0.047 ~AU$.  They call this planet the "missing link"
because of the relatively long period, which overlaps with
the planets discovered by radial velocity searches.
OGLE-TR-111-b is one of the least irradiated known transiting extrasolar planets,
(Baraffe et al. 2005, Laughlin et al. 2005), and therefore
it is also an interesting case to study because
it may probe the transition region between strongly 
irradiated and isolated planets.

We have previously carried out a selection of the most promising
OGLE planetary candidates using low dispersion spectroscopy in combination
with optical and near-infrared
photometry (Gallardo et al. 2005).  This work identified
OGLE-TR-111 as one of the most likely candidates to host exoplanets.
Gallardo et al. (2005) classify OGLE-TR-111 as a K-type main
sequence star with $T_{eff} = 4650\pm 95$ K,
located at a distance $d=850 \pm 40$ pc, with magnitudes $V=16.96$, 
$I=15.55$, and $K=14.14$, and reddening 
$E(B-V) = 0.16$. Their low dispersion spectrum shows strong Mgb band characteristic
of a metal-rich dwarf. They find that this star is intrinsically fainter ($M_V=6.82$),
and smaller ($R_s=0.71~ R_{\odot}$) than the Sun. 
Based on the high dispersion spectroscopy, Pont et al. (2004) derive similar 
stellar parameters for OGLE-TR-111: temperature $T_{eff} = 5070$ K,
gravity $log ~g = 4.8$, mass $M = 0.82 M_{\odot}$, radius $R_s = 0.85 R_{\odot}$,
and metallicity $[Fe/H]=0.12$ dex.
The stellar parameters were further improved by Santos et al. (2006),
based on high S/N spectra, deriving $T_{eff}=5044 \pm 83$, 
$log ~g = 4.51 \pm 0.36$, and $[Fe/H]=+0.19 \pm 0.07$, and assume  $R_s = 0.83 R_{\odot}$.
The values from these independent studies agree within the uncertainties. 

The known planetary parameters are in part based on the OGLE photometry.
There has been recent revisions of the radii of other confirmed OGLE planets using
high cadence, high S/N photometry with large telescopes (see Pont et al.
2006). Recently, Winn et al. (2006) presented accurate photometry of two
transits for OGLE-TR-111 in the $I$-band, 
revising the ephemeris, obtaining a period $P=4.0144479\pm 0.0000041 ~d$, and
measuring the system parameters, including 
an accurate stellar radius $R_s=0.831\pm 0.031 ~R_{\odot}$, and planet radius
$R_p=1.067\pm 0.054 ~R_J$. This planet radius is $10\%$ larger than the recent
value of Santos et al. (2006).
In this paper we present new high cadence $V$-band photometry covering a transit of 
OGLE-TR-111, giving an independent determination of the planetary radius, and
deriving an accurate period for the system. 

\section{Observations and Photometry}

The observations and photometry are described by Fern\'andez et al. (2006)
and D\'{\i}az et al. (2006).
The photometric observations were taken
with VIMOS at the Unit Telescope 4 (UT4) of the
European Southern Observatory Very Large Telescope (ESO VLT) at 
Paranal Observatory during the nights of April 9 to 12, 2005.
The VIMOS field of view consists of four CCDs, each covering 7$\times$8 arcmin,
with a separation gap of 2 arcmin, and a pixel scale of 0.205 arcsec/pixel. 
The large field of view of this instrument allows to monitor simultaneously
a number of OGLE transit candidates, in comparison with FORS at the VLT, which
has a smaller field of view (Fern\'andez et al. 2006). However,
for high precision photometry of an individual candidate FORS should be preferred
because its finer pixel scale allows better sampling (e.g. Pont et al. 2006).
Here we report on the observations of OGLE-TR-111, which was located
in one of the four monitored fields, and it happened to have a transit
during the first night of our run.

We used the Bessell $V$ filter of VIMOS, with $\lambda_0=5460{\rm \AA}$, 
$FWHM=890{\rm \AA}$. The $V$-band was chosen in order to complement
the OGLE light curves which are made with the $I$-band filter.
In addition, the $V$-band is more sensitive to the
effects of limb darkening during the transit, and is adequate for the
modeling of the transit parameters.

We have monitored two fields on April 9, 2005, one of which included the star
OGLE-TR-111. The fields were observed alternatively with three exposures
of 15s before presetting to the next field. 
For this program we managed to
reduce the observation overheads for telescope presets, instrument setups,
and the telescope active optics configuration to an absolute minimum.
This ensured adequate sampling of the transit: 
we obtained 224 points during the first night in the field of OGLE-TR-111.  The 
observations lasted for about 9.5 hours, until the field went below 3 airmasses.

In order to reduce
the analysis time of the vast dataset acquired with VIMOS, 
the images of OGLE-TR-111 analyzed here are 400$\times$400 pix, or 
80 arcsec on a side.  Each of these small images contains
about 500 stars with $15<V<24$ that can be 
used in the difference images, and light curve analysis.
The 7 best seeing images ($FWHM=0.5~ arcsec$) taken near the zenith 
were selected, and a master image was made in order to
serve as reference for the difference image
analysis (see Alard 2000, Alard \& Lupton 1998).
The candidate star is not contaminated by faint neighbours, judging
from our deep $V$ and $Ks$-band images. 
There is a $V=18.80$ mag
star 2 arcsec South of our target, which does not affect our photometry.

The difference image photometry yields a noisier light curve with a low amplitude
transit, and we decided to apply the OGLE pipeline DIA (Udalski et al. 2002).
This new reduction showed significantly reduced photometric scatter.

For OGLE-TR-111, with a mean visual magnitude $V=16.96$, we achieved a photometric
accuracy of $0.002$ -- $0.003$ magnitudes. The errors mainly depend
on the image quality, which was worse at the beginning and end of the 
time series, when the target had large airmass.

Figure 1 shows the light curve for the first night of observations, when
the OGLE-TR-111 transit was monitored. Every single datapoint is shown, no
mesurement is discarded. For comparison, Figure 1 also
shows the phased light curve of the OGLE $I$-band photometry
(in a similar scale). The transit is well sampled in the $V$-band,
and the scatter is smaller.
There are $N_t = 52$ points in our single transit shown in Figure 1, and the 
minimum is well sampled, allowing us to measure an accurate amplitude.  
In the case of OGLE, the significance of the transits is --in part-- judged
by the number of transits detected. In the case of the present study,
we compute the signal-to-noise of the single, well sampled transit.
For OGLE-TR-111 we find the S/N of this transit to be $S/N=43$  following
Gaudi (2005).
However, this does not include systematic effects (red noise),
which we consider below.

After the OGLE 2002 transit campaign, the field of
OGLE-TR-111 was observed by the OGLE survey regularly but less
frequently with the main aim of improving the ephemeris. Altogether
more than $200$ new epochs were collected in the observing seasons 2003--2005.
Unfortunately, there were no eclipses observed between 2002 and April 2005,
when OGLE started recovering eclipses. This is because
this system has a near multiple of a day period.
The full OGLE photometric dataset covering almost 350 cycles
made it possible to significantly  refine the
ephemeris of OGLE-TR-111:
$$HJD(middle ~of ~transit) = 2452330.46228 + 4.014442 * E$$
(Udalski et al. 2005, private communication).
The VIMOS transit reported here also agrees with this ephemeris, as 
discussed in Section 5.

\section{The Radius of OGLE-TR-111-b}

In light of the high dispersion
follow up of this target by Pont et al. (2004) that confirmed the low
mass of OGLE-TR-111-b, we can consider that the transit measured
here is representative of all transits of OGLE-TR-111-b. 
Unfortunately, just a single transit was measured here, and only in the $V$-band filter,
but it is of value because we used a different passband that previous works and can
therefore independently check the parameters for the system.
For example, the stellar limb darkening is different from the $I$-band,
with transits that are shallower at the edges but about $5-10\%$ deeper
in the central parts  (e.g. Claret \& Hauschildt 2003).

There are a few published measurements of the radius of the OGLE-TR-111 companion, 
based on the OGLE photometric data (Table 1).
Udalski et al. (2002) measured 
$A_I=0.019$ mag, and estimated $R_s=0.90 ~R_{\odot}$ for OGLE-TR-111, 
and a lower limit of $R_p=1.05 ~R_J$ for its companion, arguing 
this was one of the most promising extrasolar planetary transit candidates. 
Based on a fit to the same OGLE data, Silva \& Cruz (2006) estimated 
$R_p=1.16 ~R_J$ for the companion.
Pont et al. (2004) give $R_s=0.85 ~R_{\odot}$ for OGLE-TR-111,
and $R_p=1.0 ~R_J$ for its companion, also using on the OGLE data.
Gallardo et al. (2005) measure $R_s=0.71 \pm 0.02 ~R_{\odot}$ for OGLE-TR-111,
and $R_p=0.94 \pm 0.03 ~R_J$ for its companion, also based on the
OGLE amplitude.

Figure 2 shows our best fit to the transit curve, following Mandel \& Agol (2002),
using appropriate limb-darkenning coefficients for the $V$-band.
This fit yields $R_p/R_s=0.1245 ^{+0.0050}_{-0.0030}$ mag,
$a/R_s=12.11 ^{+1.00}_{-1.39}$, and $i=87-90 ~  deg$.
Using $R_s=0.83 ~R_{\odot}$ gives $a=0.0467 ^{+0.050}_{-0.065} ~AU$.
The uncertainties of the fit parameters were estimated from the
$\chi^2$ surface. As shown by Pont (2006), the existence of covariance
between the observations produces a low-frequency noise which must be
considered to obtain a realistic estimation of the uncertainties. To
model the covariance we followed Gillon et al. (2006) and obtained an
estimate of the systematic errors in our observations from the
residuals of the lightcurve. The amplitude of the white ($\sigma_w$)
and red ($\sigma_r$) noise can be obtained by solving the equation
system presented in their equations (5) and (6).
Repeating our procedure for similar VLT data on
OGLE-TR-113 (D\'{\i}az et al. 2006), we estimated the white noise amplitude,
and the low-frequency red noise amplitude,
and then, the surface which determines the uncertainty interval.
The projections of the $\chi^2$ surface to estimate 
the uncertainties of the fit parameters are shown in Figure 3,
as done by D\'{\i}az et al. (2006) for OGLE-TR-113.

The light curve was also fitted to obtain the transit time,
by fixing the other parameters (a, R$_p$ and i).
For example, the regions for $\Delta\chi^2 = 1$
(only white noise, without systematics), and $\Delta\chi^2 =
2.027$ (with white + red noise, including systematics) for the fit
parameters as function of the transit time are marked in Figure 4.

At this point, the major uncertainty on the OGLE-TR-111-b planetary 
radius arises from the uncertainties in the stellar properties.
Note that we do not fit the star radius simultaneously using the photometry
as done by Winn et al. (2006).
Adopting $R_s=0.71~ R_{\odot}$ from Gallardo et al. (2005), we obtain $R_p=0.860 ^{+0.056}_{-0.041} ~R_J$.
Adopting $R_s=0.83 ~R_{\odot}$ from Santos et al. (2006), we obtain 
$R_p=1.005 ^{+0.065}_{-0.048} ~R_J$. 
The unweighted mean of these two independent determinations
is $R_p=0.93\pm 0.04 ~R_J$.
However, we will adopt the spectroscopic determination of Santos et al.
(2006) of $R_s=0.83 ~R_{\odot}$, instead of the one from 
Gallardo et al. (2005) for two main reasons. First, the Santos spectroscopic data are
more recent and of higher quality, resulting on a complete analysis of the
composition and stellar parameters based on high dispersion spectroscopy, while Gallardo
give an indirect determination based on the surface brightness.
Second, in order to allow a direct comparison with the results of Winn et
al. (2006), who also adopt the mass from Santos et al. (2006).

Table 1 lists the previous estimates of the size for the OGLE-TR-111 transiting planet
$R_p$ from the literature, and this work, along with the stellar parameters.   
The agreement of the most recent 
values to within about $10\%$ implies that the radius of this planet is known.
The unweighted mean of the radii measured by Santos et al. (2006), 
Winn et al. (2006), and this work is:
$R_p = 1.014\pm 0.035 ~R_J.$

With this radius, OGLE-TR-111-b does not seem to be oversized for its mass, its
gross properties (mass, radius, mean density)
being similar to the Jovian planets of the Solar System, as listed
for example by Guillot (2005). 

OGLE-TR-111-b is the least irradiated of the known transiting extrasolar planets,
with equilibrium temperature $T_{eq}=900 ~K$
(Baraffe et al. 2005, Laughlin et al. 2005, Lecavelier des Etangs 2006). 
It lies at the cool end of the 
distribution of the other transiting hot Jupiters ($1000<T_{eq}<2000~ K$), 
but it is still warmer than the Solar System giants ($T_{eq}=300 ~K$).
Thus, OGLE-TR-111-b is not only a missing link regarding its orbital properties,
as suggested by Pont et al. (2004), but also may probe the transition between
strongly irradiated and more isolated planets.

It is interesting to compare OGLE-TR-111-b with HD209458b, which has a similar mass
($M_p=0.69~ M_J$), and orbital semimajor axis ($a=0.045 ~AU$). Yet the radius of HD209458b
is about 40\% larger than the radius of OGLE-TR-111-b measured here
(Laughlin et al. 2005, Baraffe et al. 2005). Two main effects could produce this large 
difference. First, HD209458b is inflated by stellar irradiation, which is
smaller for OGLE-TR-111-b. The difference in incident flux is a factor
of 4 according to Baraffe et al. (2005).
This irradiation difference is mostly because the primary star HD209458
($T_{eff}=6000~ K$), is hotter than OGLE-TR-111 ($T_{eff}=4500-5000~ K$). 

Second, the presence of a massive solid core in OGLE-TR-111-b
might make its radius smaller  in comparison with HD209458b.
Models predict that
the presence of a massive solid core should reduce the radius of a giant
planet significantly (Saumon et al. 1996, Burrows et al. 2003,
Bodenheimer et al. 2003, Sato et al. 2006, Guillot et al. 2006). For example,
the reduction in radius is $10$\% for a $1M_J$ planet with a 
$50M_E$ core with respect to a planet with a small core. 

\section{Times of Transit and the Orbital Period}

It has been recently realized that it is very important to measure the transit
times accurately, because of the exciting possibility of using these times 
for future studies of multiplicity in these systems
(Holman \& Murray 2005, Agol et al. 2005).

Winn et al. (2006) measure two transits accurately, giving an improved period of:
$P = 4.0144479 \pm 0.0000041 ~days.$
The mean transit times  measured by Winn et al. (2006) are:
$T_c[HJD]= 2453787.70854 \pm 0.00035,$ and
$T_c[HJD]= 2453799.75138 \pm 0.00030.$
These transits are separated by 3 cycles, and we use their mean along with
our own observations to compute a more accurate period.
The VLT transit occurred  80.5 cycles before the mean of the transits observed by 
Winn et al. (2006). 
The mean transit time  measured at the VLT is:
$T_c[HJD]= 2453470.56397 \pm 0.00076,$
and the final period measured here is:
$P=4.014484 \pm 0.000014 ~days.$
This period is independent of the OGLE photometry, but it is consistent with the OGLE data.
The errors include the systematic errors (Figure 4).

Therefore, our improved ephemeris for the mean transit times of OGLE-TR-111-b is:
$$HJD(middle ~of ~transit) = 2453470.56397 (76) + 4.014484 (14) * E,$$
where the numbers in parenthesis indicate the errors in the last two digits of their
respective quantities.

Adopting the ephemeris of Winn et al. (2006), the mean predicted time
is off from the center of our transit, which occurs about 5 minutes earlier.
In fact, the period determined here is more than $8.8\sigma$ away from the period measured by
Winn et al. (2006), or $2.6\sigma$ away using our larger errorbars (that include systematics). 
Nonetheless, we believe that it is accurate, because it relies
on their two well sampled transits as well as our single transit. 

Interestingly, there is a difference with the previous measured periods.
Such difference can arise in the presence of another massive planet in the system,
which is the main motivation for measuring accurate timing of this planet. 
With the present data, however, it cannot be claimed that this is the case:
we cannot yet rule out a constant period.
More transits should be accurately measured in the following seasons
to follow up this interesting system and confirm or rule out variations
in the mean transit times.

Due to the period very close to $4 ~d$, about 20 consecutive full transits
can be observed during a season for a period of three months, 
and then they are not observable again for a period of about six months.
With the new ephemeris we are able to predict that the next observable
series of transits of OGLE-TR-111 occur around December 2006, and then again in mid-2008,
as shown in Figure 5.  It is evident that the window of opportunity for
accurate photometric measurements of the transits for this target is small.
An alternative way to track another massive planet in the system is to perform
radial velocity follow-up with no critical time of observational windows,
even though this method is time consuming and difficult for such a faint candidate.
 
\section{Conclusions}

Udalski et al. (2002) discovered low amplitude transits in the main sequence
star OGLE-TR-111, which we observed with VIMOS at the ESO VLT.
The planet OGLE-TR-111-b is a massive planet with mass $M_p = 0.53 ~M_J$ 
(Pont et al. 2004).  
We were able to accurately sample one low amplitude transit on this star
with high cadence observations, complementing the recent measurement of two 
transits by Winn et al. (2006).

We improve upon the parameters of the transit, in particular measuring
the transit time, $t_T=2.9\pm 0.1$ hours, the orbital inclination
$i=86.9 - 90 ~degrees$, and the orbital semimajor axis $a=0.047 ^{+0.03}_{-0.06} ~AU$. 
These data are  in agreement with the orbital parameters measured by
radial velocities and with the stellar parameters.

The two main results of this work are:

(1) We measure an accurate radius based on VLT transit photometry and revised
stellar parameters.
We obtain $R_p=1.005 ^{+0.065}_{-0.048} ~R_J$, which for a mass of
$M_P=0.53 ~M_J$, gives a density of $\rho_p=0.6\pm 0.2 ~g/cm^3$.
Thus, the planet of OGLE-TR-111 has a radius similar 
to Jupiter, and a mean density that resembles that of Saturn.
The OGLE-TR-111-b planet does not appear to 
be significantly inflated by stellar radiation like HD209458-b.

(2) Using newly available transits from Winn et al. (2006), we are able to 
measure accurately the orbital period of this interesting system, 
$P=4.014484 \pm 0.000014 ~days,$
as well as to update the ephemeris.
The timing is different from previously published values, but there is not
yet sufficient data to claim time variations caused by another massive planet 
in the system. Follow up of the OGLE-TR-111-b transits is warranted in the
next observable transit windows around December 2006 and mid-2008.

\acknowledgments{
DM, JMF, GP, MZ, MTR, WG are supported by Fondap Center for Astrophysics 
No. 15010003.  AU acknowledges support
from the Polish MNSW DST grant to the Warsaw University Observatory.
We also thank the ESO staff at Paranal Observatory.
}

\clearpage

\begin{figure}
\plotone{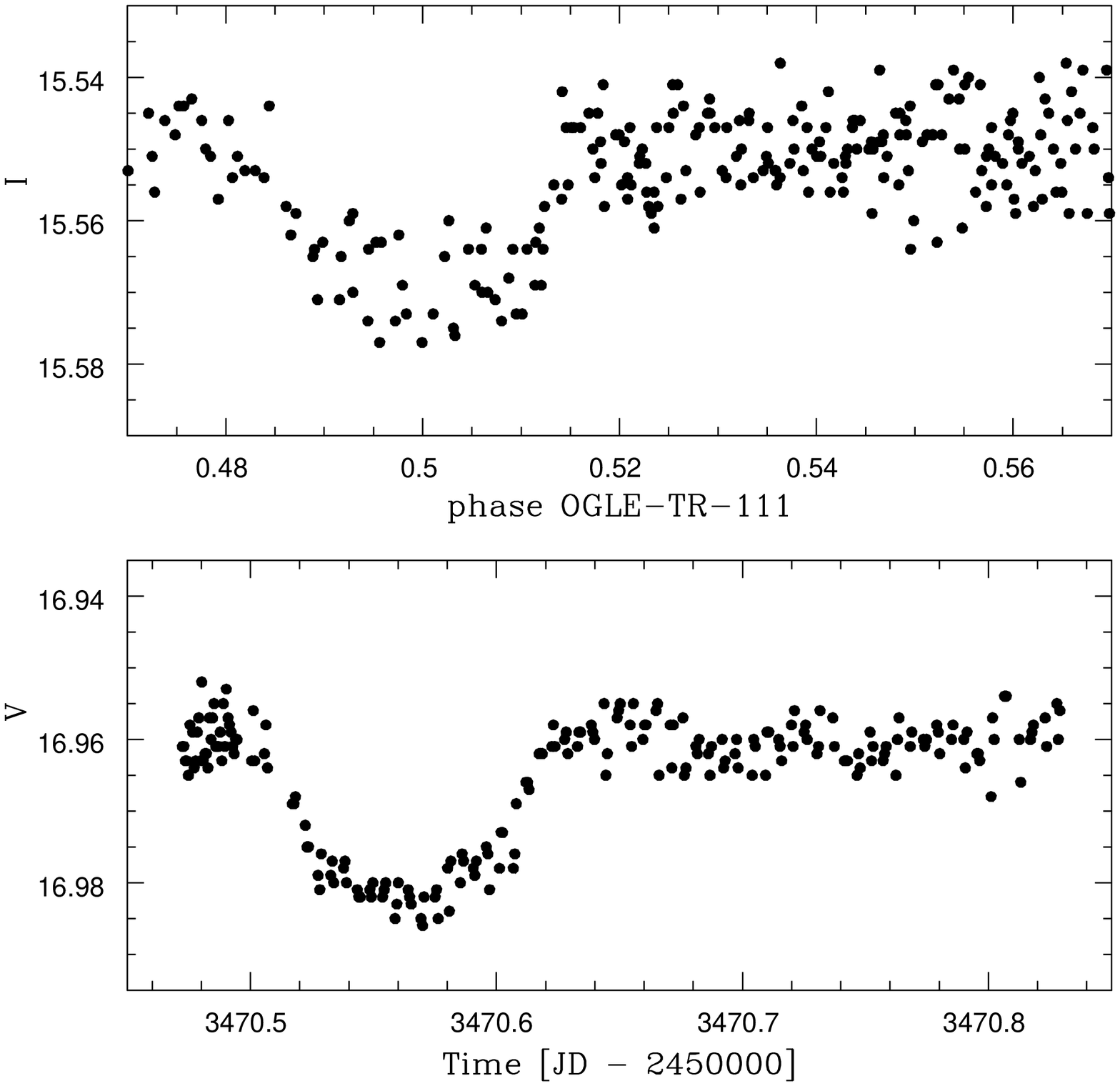}
\caption{
Single transit of OGLE-TR-111 observed during the first night (April 9, 2005) with VIMOS
in the $V$-band (bottom) compared with the OGLE phased light curve transit
in the $I$-band (top). The star is $1.5$ mag fainter in the $V$-band, but
in spite of this the smaller scatter of the VIMOS photometry is evident.
}
\end{figure}

\clearpage

\begin{figure}
\plotone{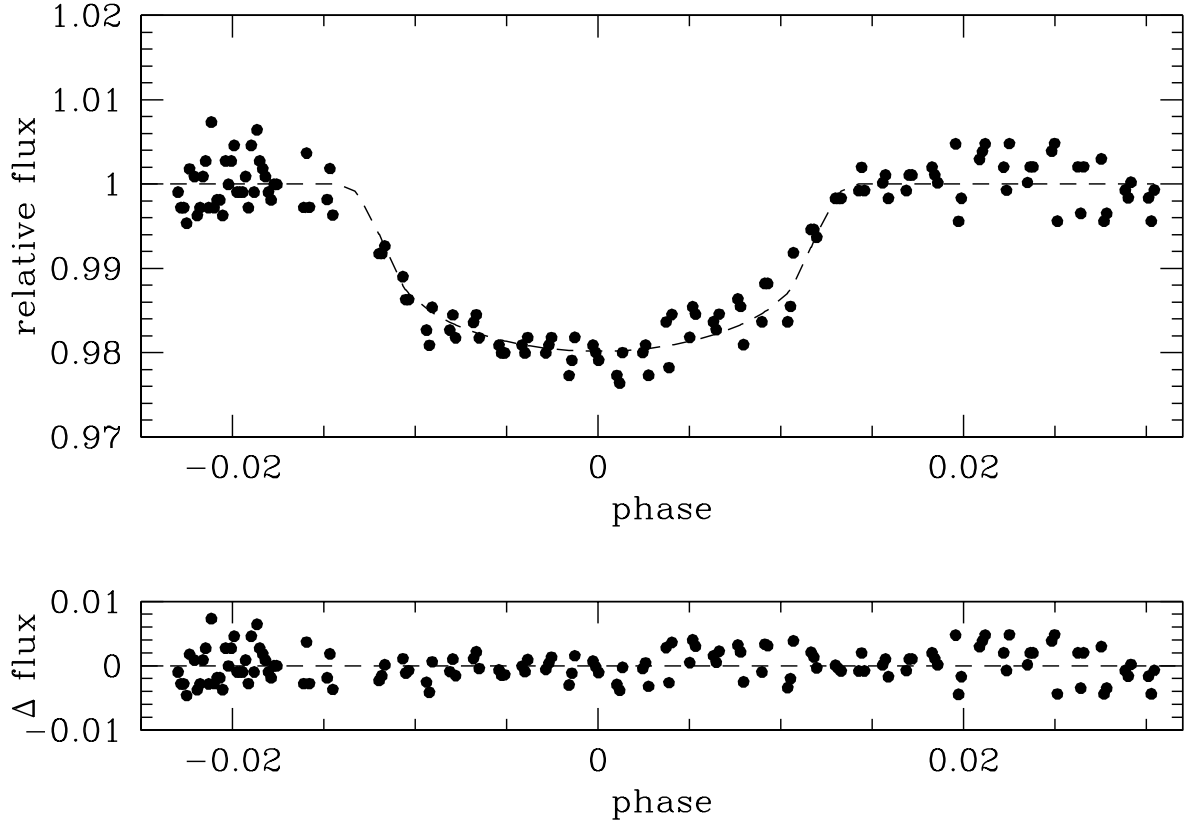}
\caption{
Fit to the single transit of OGLE-TR-111 observed  in the $V$-band
during the first night of the VIMOS run (top), and the residuals (bottom). 
}
\end{figure}

\clearpage

\begin{figure}
\plotone{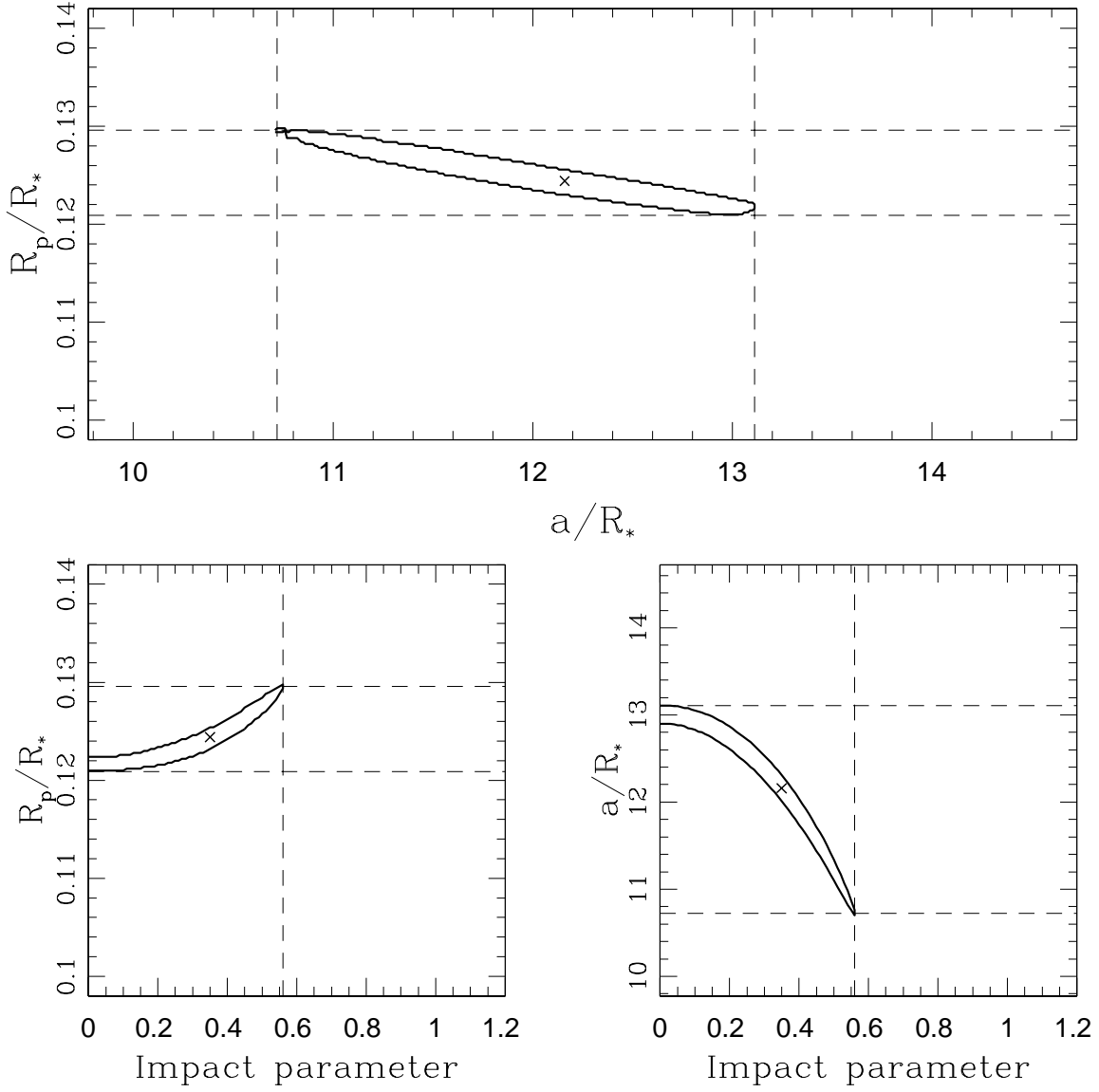}
\caption{
Regions with $\Delta\chi^2 = 2.027$ (with white + red noise, including systematics)
for three projections of the fit parameters.  The regions with $\Delta\chi^2 < 1.0$
(only white noise, without systematics), differ only slightly and are not plotted 
to avoid confusion.
}
\end{figure}

\clearpage

\begin{figure}
\plotone{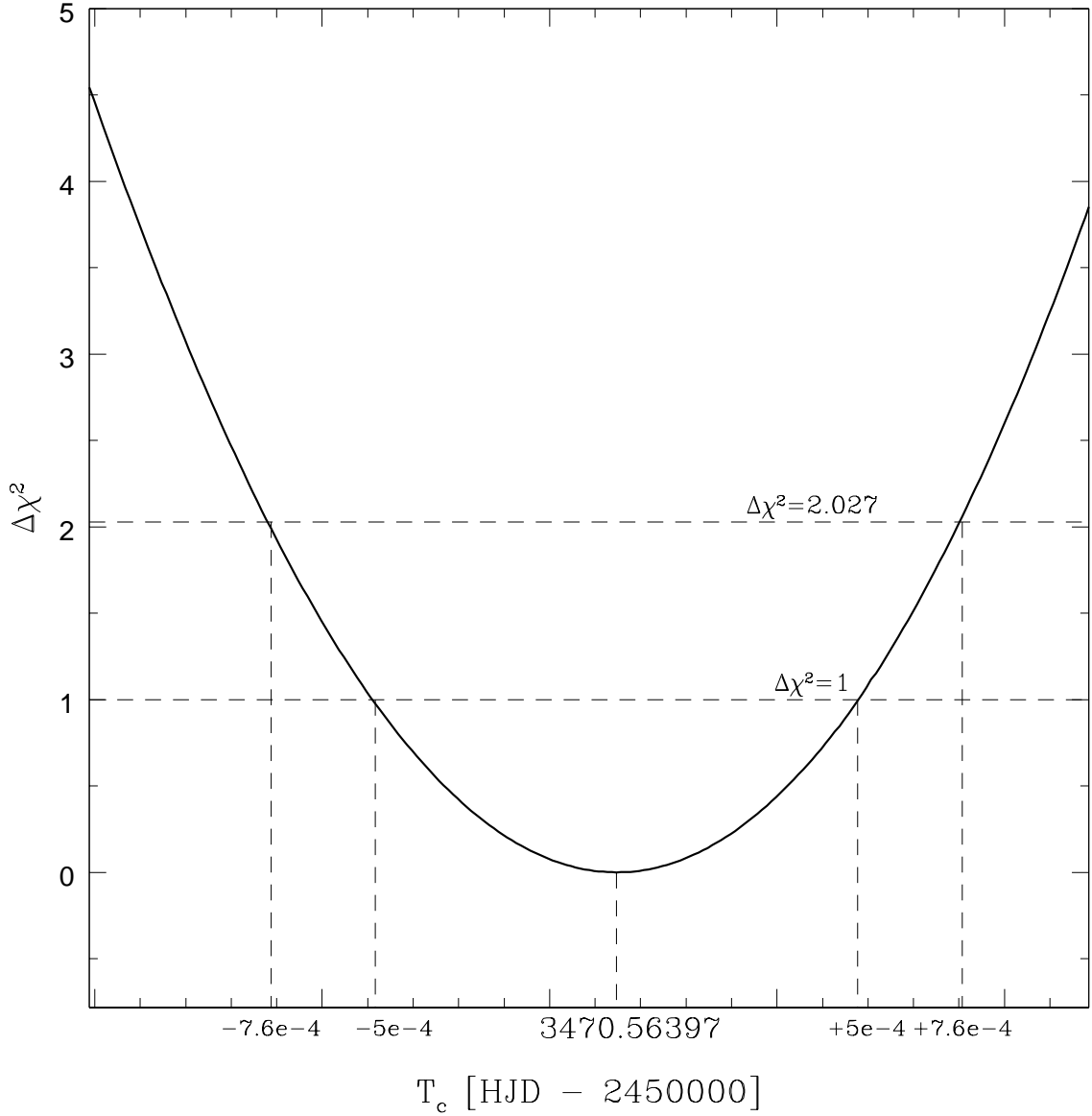}
\caption{
$\Delta\chi^2$ as function of the transit time obtained by fitting the transit curve.
The horizontal dashed lines show $\Delta\chi^2 = 1.0$
(only white noise, without systematics),
and $\Delta\chi^2 = 2.027$ (with white + red noise, including systematics).
}
\end{figure}

\clearpage

\begin{figure}
\plotone{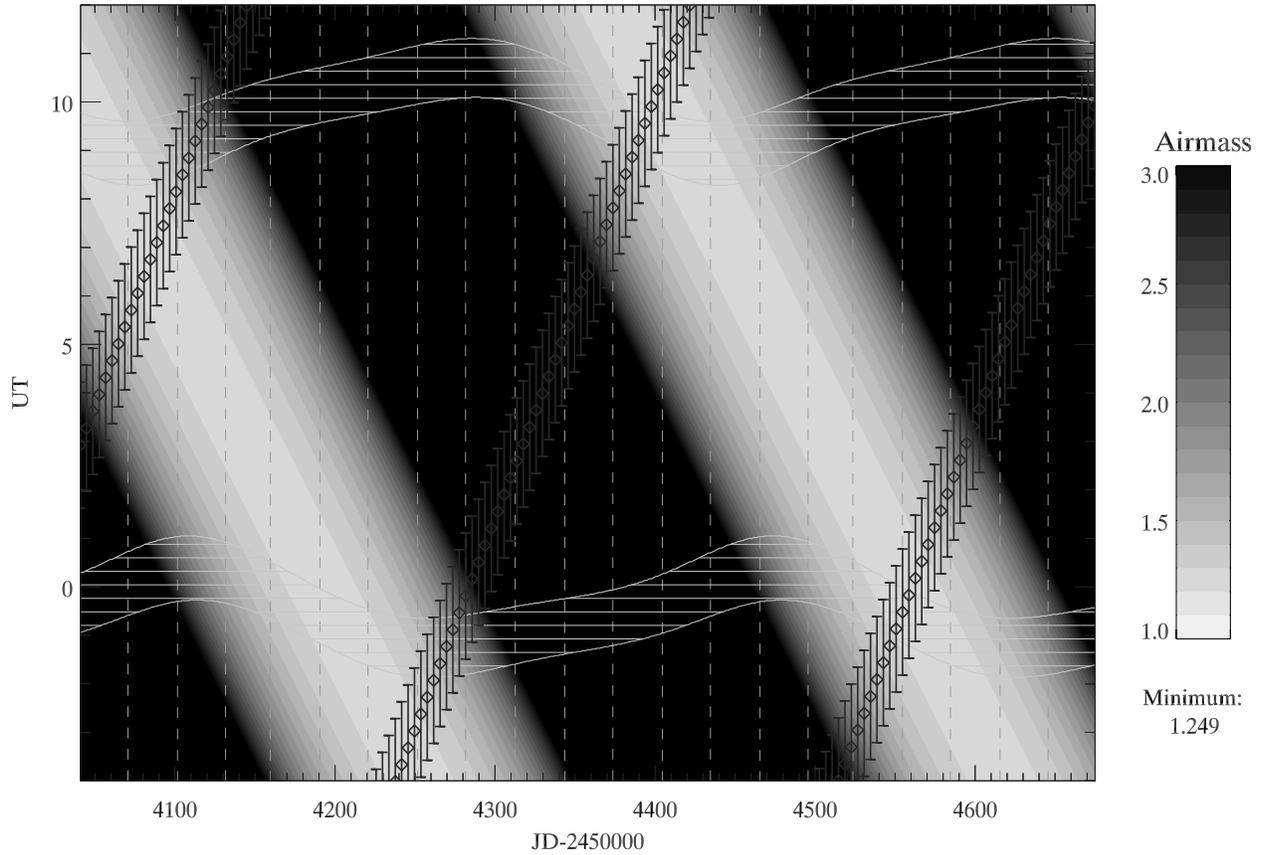}
\caption{
Observable series of transits for OGLE-TR-111 from November 2006 until August 2008, 
as seen from the telescopes in Northern Chile (latitude 24$^o$ 40' South).
The diamonds indicate the
center of the transit events; their vertical error bars indicate transit
span.   Line-filled regions indicate morning and evening twilight,
respectively, and dashed vertical lines indicate month boundaries.
The increasing shades indicate increasing airmass, as shown in the scale to the right. 
}
\end{figure}

\clearpage

\begin{deluxetable}{llllll}
\tablewidth{0pt}
\tablecaption{Measurements of the OGLE-TR-111-b Radius}
\tablehead{
\multicolumn{1}{c}{Reference}&
\multicolumn{1}{c}{$M_s [M_{\odot}]$}&
\multicolumn{1}{c}{$R_s [R_{\odot}]$}&
\multicolumn{1}{c}{$R_p [R_J]$}&
\multicolumn{1}{c}{Comments}}
\startdata
Udalski + 2002 &$1.0$ &$0.9$ &$1.05$         & lower limit, 9 transits, $I$-band phot.\nl
Pont + 2004    &$0.82 ^{+0.15}_{-0.02}$ &$0.85$ &$1.00\pm 0.10$ & using OGLE phot., new stellar par.\nl
Gallardo + 2005&--- &$0.71$ &$0.94\pm 0.03$ & using OGLE amp., new stellar par.\nl
Silva + 2006  &$0.96\pm 0.15$ &--- &$1.16\pm 0.19$ & re-analysis of OGLE phot.\nl
Santos + 2006  &$0.82\pm 0.02$ &$0.83$ &$0.97\pm 0.06$ & using OGLE phot., new stellar par.\nl
Winn + 2006    &$0.82\pm 0.02$ &$0.83\pm 0.03$ &$1.067\pm 0.054$ & two transits, $I$-band phot.\nl
This work             &$0.82\pm 0.02$ &$0.83\pm 0.03$ &$1.005 ^{+0.065}_{-0.048}$ & single transit in $V$-band\nl
\enddata
\end{deluxetable}


\begin{references}
\reference{}{Agol, E., et al. 2005, MNRAS, 359, 567 }
\reference{}{Alard, C. 2000, A\&ASuppl., 144, 363 }
\reference{}{Alard, C., \& Lupton, J. 1998, ApJ, 503,325 }
\reference{}{Baraffe, I., Chabrier, G., Barman, T., Selsis, F., Allard, F., \& Hauschildt, P. H. 2005, A\&A, 436, L47}
\reference{}{Bodenheimer, P., Laughlin, G., \& Lin, D. N. C. 2003, ApJ, 592, 555 }
\reference{}{Burrows, A., Sudarsky, D., \& Hubbard, W. B. 2003, ApJ, 594, 545 }
\reference{}{Claret, A.,  \& Hauschildt, P. 2003, A\&A, 412, 241 }
\reference{}{D\'{\i}az, R., et al. 2006, ApJ, submitted}
\reference{}{Fern\'andez, J. M., et al. 2006, ApJ, 647, 587}
\reference{}{Fischer, D. A., et al. 2005, ApJ, 620, 481 }
\reference{}{Gallardo, J., Minniti, D., Valls-Gabaud, D., \& Rejkuba, M. 2005, A\&A, 431, 707}
\reference{}{Gaudi, S. A. 2005, ApJ, 628, L73 }
\reference{}{Gillon, M., Pont, F., Moutou, C., Bouchy, F., Courbin, F., Sohy, S., \& Magain, P. 2006, A\&A, 459, 249 }
\reference{}{Guillot, T. 2005, Ann. Rev. Earth Planet Sci., 33, 493 }
\reference{}{Guillot, T., et al. 2006, A\&A, 453, L21 }
\reference{}{Holman, M., \& Murray, N.,  2005, Science, 307, 1288 }
\reference{}{Lecavelier des Etangs, 2006, A\&A, in press (astroph/0609744)}
\reference{}{Laughlin, G., et al. 2005, ApJ, 621, 1072 }
\reference{}{Mandel, K., \& Agol, E. 2002, ApJ, 580, L171}
\reference{}{Miralda-Escude, J. 2002, ApJ, 564, 1019}
\reference{}{Pont, F., Bouchy, F., Queloz, D., Santos, N. C., Melo, C., Mayor, M., \& Udry, S.  2004, A\&A, 426, L15 }
\reference{}{Pont, F., et al.  2006, A\&A, submitted (astro-ph/0610827) }
\reference{}{Pont, F., Zucker, S., \& Queloz, D.  2006, MNRAS, 373, 231 }
\reference{}{Sahu, K. 2006, et al. Nature, 443, 534}
\reference{}{Santos, N., et al.  2006, A\&A, 450, 825 }
\reference{}{Sartoretti, P., \& Schneider, J. 1999, A\&ASuppl, 134, 553 }
\reference{}{Sato, B., et al. 2005, ApJ, 633, 465 }
\reference{}{Saumon, D., Hubbard, W. B., Burrows, A., Guillot, T., Lunine, J. I., \& Chabrier, G. 1996, ApJ, 460, 993}
\reference{}{Silva, A. V. R., \& Cruz, P. C. 2006, ApJ, 642, 458}
\reference{}{Udalski, A., et al. 2002, Acta Astronomica, 52, 317}
\reference{}{Winn, J. N., Holman, M. J., \& Fuentes, C. I. 2006, AJ, in press (astro-ph/0609471) }
\end{references}
\end{document}